# Localizing axial dense emitters based on single-helix point spread function and compressed sensing

HANZHE WU,[1,3] DANNI CHEN,[1,3,*] YIHONG JI,[1] GAN XIANG,[1] HENG LI,[2] BIN YU,[1] JUNLE QU[1]

*1 College of Physics and Optoelectronic Engineering, Key Laboratory of Optoelectronic Devices and Systems of Ministry of Education and Guangdong Province, Shenzhen University, Shenzhen 518060, China*
*2 Tsinghua-Berkeley Shenzhen Institute (TBSI), Tsinghua University, Shenzhen, 518055, China*
*3 contributed equally to this work*
*\*danny@szu.edu.cn*

**Abstract:** Among the approaches in three-dimensional (3D) single molecule localization microscopy, there are several point spread function (PSF) engineering approaches, in which depth information of molecules is encoded in 2D images. Usually,the molecules are excited sparsely in each raw image. The consequence is that the temporal resolution has to be sacrificed. In order to improve temporal resolution and ensure localization accuracy, we propose a method, SH-CS, based on light needle excitation, detection system with single helix-point spread function (SH-PSF), and compressed sensing (CS). Although the SH-CS method still has a limitation about the molecule density, it is suited for relatively dense molecules. For each light needle scanning position, an SH image of excited molecules is processed with CS algorithm to decode their axial information. Simulations demonstrated, for random distributed 1 ~ 15 molecules in depth range of 4 μm, the axial localization accuracy is 12.1 nm ~ 73.5 nm. The feasibility of this method is validated by experimental data.

## 1. Introduction

In recent years, several approaches have been put forward in the field of super-resolution microscopy. As one of the common methods, 2D SMLM's performance is outstanding in spatial resolution which is usually less than 20 nm. In order to study 3D samples, 3D SMLM was developed, where other technologies with axial localization ability were introduced into SMLM, including astigmatic localization [1], PSF engineering [2, 3], bifocal plane detection [4], fluorescence interferometry [5], etc. All of the SMLM approaches work on a premise that excited molecules in each raw image should be sparse, so they are very dependent on the photoswitchable characteristic of fluorescent dyes [6]. Furthermore, the temporal resolution of these methods is quite limited.

Compared with the SMLM where molecules are excited sparsely by taking advantage of photoswitchable fluorescent dyes, there is another excitation strategy of reducing the density of effectively excited molecules, i.e. reducing the full width at half maximum (FWHM) of the effective excitation PSF. At the same time, reduced FWHM means better spatial resolution. STED is one representative method with the second strategy. By overlapping a donut-shaped depletion beam on a Gaussian-shaped excitation beam, the stimulated fluorescence molecules in the non-central region return to the ground state by stimulated radiation with certain probabilities, thus reducing the effective fluorescence emission PSF. Since the high intensity of depletion beam implies a smaller point PSF, its lateral resolution can even reach several nanometers [7, 8]. In order to maintain a good lateral resolution at a large imaging depth, a Gauss-Bessel STED (GB-STED) microscopy was subsequently proposed, in which a Gaussian and a higher-order Bessel beam were used as the excitation and the depletion beam respectively [9]. However, due to the confocal detection adopted in GB-STED, time-consuming 3D scanning is required to get the 3D information of samples. Subsequently, researchers proposed that both the excitation beam and the depletion beam could adopt a non-diffracted Bessel beam, BB-STED, so samples could be effectively excited with a light needle whose FWHM can be tens of nanometers and depth of field can be several microns [10, 11]. So far, the light needle has been used in light-sheet microscopy to increase the field of view and axial resolution [12]. In BB-STED, although the lateral density of excited molecules could be reduced by BB-STED effectively, the axial density remains the same. If the axial information of the excited molecules in the light needle can be recovered, then the volumetric images of the sample can be obtained by only a 2D scanning. To achieve this, a method should be developed to locate the axially dense molecules in a light needle simultaneously.

Actually, there are a variety of algorithms of dense molecules localization, which mainly belong to two categories. Algorithms in the first category were developed based on the sparse molecular localization algorithms. Representative algorithms are DAOSTORM (Three Dimension Dominion Astrophysical Observatory Stochastic Optical Reconstruction Microscopy) [13], SSM_BIC (Structured Sparse Model_Bayesian Information Criterion) [14]. Algorithms in the other category focus on estimating the molecular density to obtain the maximum probability of the density distribution. Representative algorithms are Compressed Sensing (CS) [15], and 3B (Bayesian analysis of the Blinking and Bleaching) [18]. As the most representative algorithm in the first category, DAOSTORM achieved a lateral localization accuracy of 20 nm when the lateral projection density was less than 1 $\mu m^{-2}$. However, when the density exceeded 1 $\mu m^{-2}$, the localization accuracy decreases significantly. SSM_BIC employs the sparse structure model to estimate the initial multi-molecule model, and then selects the optimal fitting model in conjunction with Bayesian information criteria. Therefore, it can achieve sufficiently high localization accuracy and recover rate under the condition of low SNR (Signal-to-Noise Ratio), but execution speed is very slow. Compared with the algorithms in first category, those in the second one can handle situations with higher molecule density. For example, when CS was introduced into STORM, for images with lateral density even up to 12 $\mu m^{-2}$, the lateral localization accuracy could still reach 10 ~ 60 nm. Furthermore, CS was also used to localize dense molecules in three dimensions [16, 17]. Later, Anthony Barsic proposed using Double-Helix PSF (DH-PSF) combined with Matching Pursuit (MP) algorithm for rough localizing. And then Convex Optimization (CO) algorithm was used for accurate localization. Density limit of this method achieved 1.5 $\mu m^{-2}$ with an axial localization accuracy of ~ 75 nm, the depth of field of the system was only 2 $\mu m$, but the computation took a long time because of complicated procedure.

Therefore, we propose a method abbreviated as SH-CS which combines SH-PSF and CS to realize localization of axial dense molecules under light needle excitation. Double helix PSF (DH-PSF) [19] and SH-PSF [20] are both spiral PSFs with ability of spreading spots of emitters at different depths to different azimuths. The reason SH-PSF is used here instead of DH-PSF is that SH-PSF's energy is more concentrated, the SNR is better, and the depth of field is greater. After axial dense molecules are imaged with a detection system with SH-PSF, the highly overlapping spots in the image are then localized using CS. Simulation results demonstrated that for emitters density less than 15 in a depth range of 4 $\mu m$, the axial localization accuracy reaches 12.1 ~ 73.5 nm. In combination with 50 nm light needle scanning, 3D structure of the sample is well reconstructed with 3D super-resolution. The performance of SH-CS is also validated with experimental data.

## 2. Method

The principle of SH-CS is shown in Fig. 1. Under light needle excitation, the excited emitters from different depths are detected by a detection system with SH-PSF. The depth information of excited emitters is reconstructed using CS algorithm.

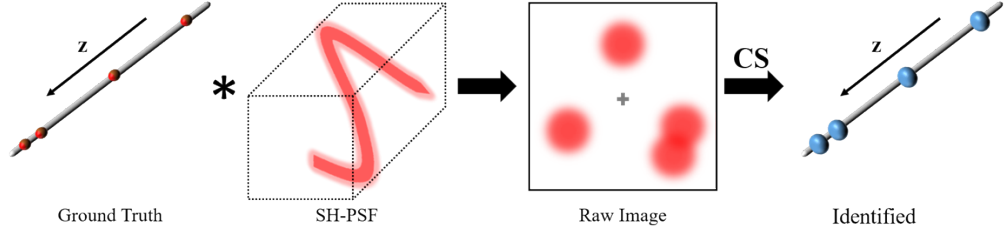

Fig. 1. The principle of SH-CS.

The non-diffraction light nano-needle could be realized by BB-STED where the concentric 0th-order Bessel beam and 1st-order Bessel beam are used as excitation and depletion beam respectively [10, 11]. By adjusting their powers, the non-diffractive light nano-needle excitation of sample can be realized. The 3D distribution of emitter emission is $I(x,y,z) = I_n(x,y,z) \cdot S(x, y, z)$, where $I_n(x,y,z)$ is the intensity distribution, and $S(x, y, z)$ represents the emitter distribution in the sample. The 3D distribution of fluorescence is $I_{img}(x, y,z) = I(x,y,z) * H(x,y,z)$, where $H(x,y,z)$ is the SH-PSF of the detection system. The image of an emitter acquired by an array detector is a spot at a certain azimuth with its lateral coordinate as the origin, and the final image $y$ is an overlap of all spots at several azimuths corresponding to emitters at different depths.

The lateral coordinates of emitters in the reconstruction result are the center of light needle, and their axial coordinates are recalled by CS algorithm. After the sample is excited by a light needle, the axial distribution of excited emitter is a vector $x$ to be sought, and the acquired image is taken as vector y which can be calculated as

$$\boldsymbol{y = Ax} \tag{1}$$

where $A$ is the measurement matrix which could be build based on the 3D-PSF of the system.

The construction process of the measurement matrix $A$ is shown in Fig. 2. First, the effective axial range is equally divided into $s$ layers (Fig. 2(a)). Next, as is shown in Fig. 2(b), the matrix $M_i$ ($m \times m$) corresponding to SH-PSF image at the depth of layer $i^{th}$ is transformed into the $i^{th}$ column of measurement matrix $A$ by connecting the columns of matrix $M_i$ end to end. Eventually a measurement matrix $A$ with $m^2$ rows and $s$ columns is constituted. The optimal solution $x$ of Equation (1) is solved according to the constraint condition.

$$\boldsymbol{min\ ||x||_1\ subject\ to\ ||Ax - y||_2 \leq \varepsilon \cdot \sqrt{\sum y_i}} \tag{2}$$

where $\varepsilon$ is an empirical value, typically ranging from approximately 0.22 to 0.3. The solved vector $x$ is the depth information reconstructed from the acquired image $y$.

The final 3D structure is achieved by scanning in the *xoy* plane with a light needle.

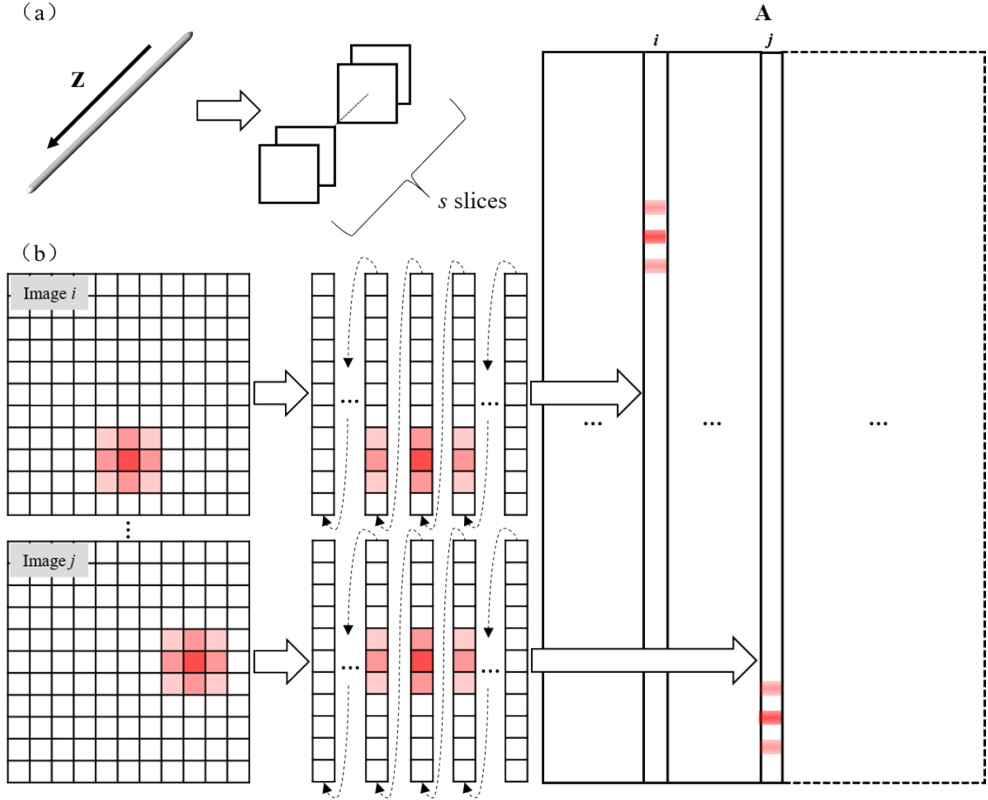

Fig. 2. Schematic of CS in SH-CS. (a) The principle of axial division of measurement matrix *A* in SH-CS; (b) The different columns of measurement matrix *A* are constructed from SH-PSF for different depths.

## 3. Simulation

First, the axial localization capability of SH-CS is tested through simulation. The simulated detection optical path is shown in Fig. 3(a). The fluorescence signal (wavelength 560 nm) from excited emitters is collected by an objective lens (NA1.4, 100×), imaged by a tube lens (focal length 180 mm) to the first image plane, and then pass through a 4*f* relay system. The PSF of the detection system is modulated into SH-PSF by placing a special phase pattern on its Fourier plane. The phase pattern is designed according to the reference [20]. Modulated PSF of the system is shown in Fig. 3(c). Spots center at different azimuths in 2D plane with different axial positions.

According to the method introduced in [18], different number of Fresnel zones *L* will lead to different modulation depth of SH-PSF. As is shown in Fig. 3(b), more Fresnel zones (larger *L*) mean larger modulation depth. Theoretically, the value of *L* can approach infinite. However, with the increase of *L*, the effective size of PSF image obtained after modulation also increases (as shown in Fig. 3(c)), which will cause the longer execute time of CS algorithm. Therefore, $L = 9$ is subsequently adopted to balance the modulation depth and execute time. In addition, since the azimuths of 0° and 360° is practically indistinguishable, the effective angle range of phase with $L = 9$ is set to 300°, corresponding to an effective depth range of 4 μm. The axial density of emitters is represented by the number of spots in azimuth range of 300°, i.e. the number of depths in the depth range of 4 μm, which is denoted by *n*. The effective pixel size of the detector is set to 80 nm, and placed on the rear focal plane of Lens2. Gaussian noise (mean 0, variance 0.01) and Poisson noise are added to each raw image.

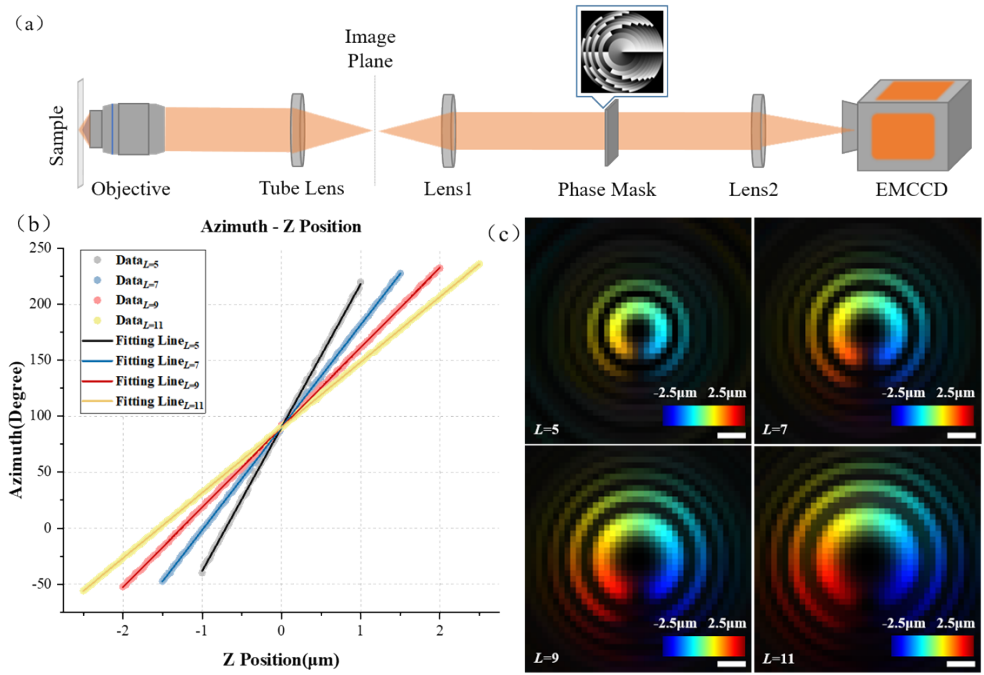

Fig. 3. Setup and PSFs of SH-CS. (a) Setup of the detection optical path; (b) The relationship between azimuth Angle and Z-axis position; (c) z-axis projection of PSF with different *L*.
Scale bar: 400 nm

First, localization performance of the SH-CS is analyzed with simulated samples in which emitters were located at different axial positions and a fixed lateral position. For a certain axial density $n$, 2000 random samples were generated, and each sample consists of $n$ emitters at random $n$ depths. Corresponding 2000 original images were simulated for the 2000 random samples. Take $n = 8$ as an example to illustrate the analysis process. Fig. 4(a) shows one of the original images. As is mentioned in section '2. Method', measurement matrix $A$ is crucial for the final localization accuracy, and the number of columns, $s$, determines the theoretical upper limit of localizing accuracy in $z$-axis. So, the original image (Fig. 4(a)) was analyzed when $s$ is set to 41, 81, 161, 201. The localization results (Fig. 4(b)) demonstrated that, depth positions of the 8 emitters are reconstructed under different $s$, but there are certain localizing deviations from the ground truth (GT) positions. Next, referring to the method of Hugelier [21], we measured localizing bias, with a pre-defined tolerance of ± 200 nm. As is shown in Fig. 4(c), for the depth $d$ where identified molecule is located, if there are GT molecules existing within the depth range of $d ± 200$ nm, the identified molecule and the GT molecule are considered to be successfully matched. If there are two identified molecules that correspond to one GT molecule at the same time, the midpoints of the two identified molecules are taken as one virtual molecule, and the virtual molecule is considered to match the GT one. For $n = 8$ and $s = 201$, 7.8 molecules were matched averagely, corresponding to 'identified number' of 7.8 at 'excited number' of 8 in Fig. 4(e). In addition, the axial deviation (denoted by $Z_D$) between matched identified molecule and GT molecule is statistically in a histogram (Fig. 4(d)) and the corresponding standard deviation (denoted by '$Z_D$ Stdev') represents the axial localizing accuracy at 'excited number' of 8 in Fig. 4(f). Average executei time for each image under different $n$ is shown in Fig. 4(g). It should be noted that, generally, with a smaller tolerance size, a method will exhibit lower recall rate but higher accuracy. So, the performance of a method should be evaluated comprehensively based on the two parameters.

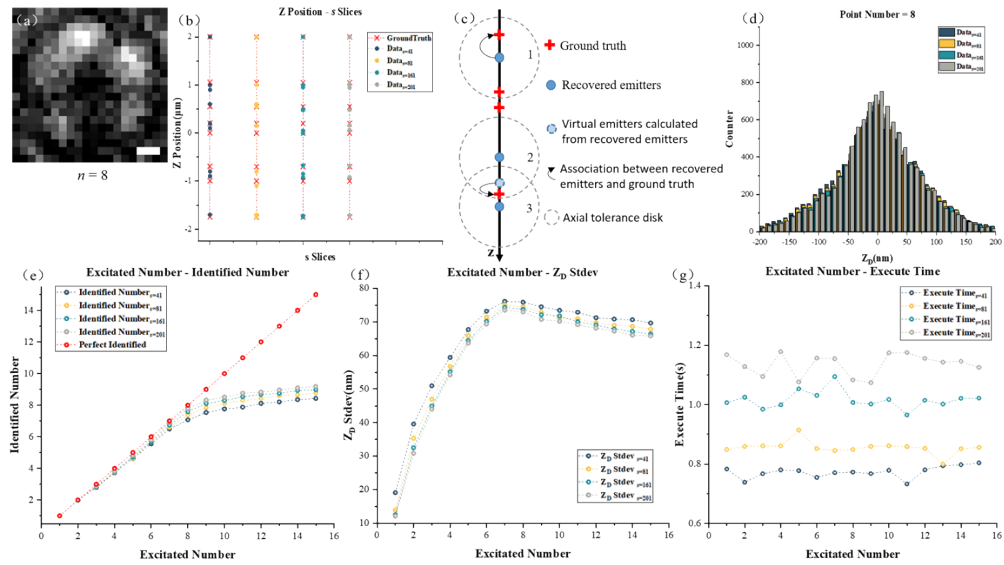

Fig. 4. Evaluation of localization performance of SH-CS. (a) Original image of SH-CS at $n = 8$, scale bar: 240 nm; (b) Localizing performance under different s with $n = 8$; (c) Measurement criteria for axial localizing; (d) Statistical histogram of axial deviation with $n = 8$; (e) Relationship between identification density and $n$; (f) Relationship between localizing accuracy and $n$; (g) Relationship between single frame execute time and $n$.

As is shown in Fig. 4(e), the number of matches increases with $n$, but recall rate which is calculated with 'identified number' / $n$ gradually decreased. However, at $n = 15$, the recall rate remains to be 61.2% ($s = 201$). In addition, under the same $n$, identified number increases monotonically with $s$. Take '$n = 10$' as an example, when $s$ is small relatively but increases from 41 to 81, identified number increases from 7.8 to 8.1, i.e., an increase of 3.8%, while for larger $s$, when a same increment of 40 of $s$ from 161 to 201, identified number increases by 2.4% only, from 8.3 to 8.5. The results demonstrated that for larger $s$, the ability of improving identified number by increasing $s$ becomes more and more limited. Fig. 4(f) represents the relationship between 'localizing accuracy' and $n$. With the increase of $n$, localizing accuracy gradually decreases, but for $n$ larger than 7, localizing accuracy seems to increases abnormally. The reason for this seemingly counterintuitive phenomenon is that when $n$ is too high, a GT molecule will easily be found near an identified molecule, which will be considered as matching. Fig. 4(f) also shows that with the increase of $s$, the localizing accuracy shows a trend of improvement. Taking $n = 6$ as an example, the localizing accuracy increases from 73.2 nm to 71.4 nm, an increase of 2.5%, when $s$ increases from 41 to 81. For bigger $s$, the same increment of 40 layers, i.e., $s$ increases from 161 to 201, brings an increase of only 1.1%, from 70.2 nm to 69.4 nm. It shows that further increasing $s$ has more and more limited ability to improve localizing accuracy. For $s = 201$, the axial localizing accuracy is within the range of 12.1 nm-73.5 nm.

Since CS is used in SH-CS, execute time of the algorithm is a concern. It should be noted that the SH-CS method adopts strategy of light needle excitation, the excited area is very constrained, and the size of raw images to be processed is only 20×20 pixel$^2$. Fig. 4(g) represents the relationship between average frame execute time and $n$. The execute time did not increase significantly with the increase of $n$. Larger $s$, means longer execute time. Even so, when $s = 201$, average execute time for a raw image is only 1.1 s ($n = 15$). Based on the above analysis, $s$ is set to 201 to maximize the reconstruction performance in the data processing thereafter.

Next, a sample with size of 2 μm × 2 μm × 600 nm was simulated, which consists of five letter-shaped structures, 'H', 'I', 'S', 'Z' and 'U'. The five letters are on five different depths,

$d_i$ ($i = 1 \sim 5$), with a gap of 100 nm between adjacent layers. The diagram of the sample's 3D structure was shown in Fig. 5(a), and in order to show the sample structure clearly, the lateral and axial coordinates take different scales bars. The structures of the five layers are shown in Fig. 5(b). The sample was scanned on the plane $xoy$ by a light needle with a 2D Gaussian distribution (FWHM = 50 nm) at a step size of 20 nm. The SH raw image acquired at each light needle excitation is analyzed by the above CS method, and the reconstructed axial structure is reconstructed as the structure at lateral coordinates corresponding to the center of the light needle. The reconstructed structures within depth ranges of $d_i \pm 50$ nm are extracted and shown in Fig. 5(c). The reason of using data in a thickness of 100 nm instead of fixed depths $d_i$ is to match the axial localizing accuracy of SH-CS. As is shown in Fig.4f, the axial localizing accuracy varies according to axial density and in a range between 12.1 nm and 73.5 nm, so the middle value of 42.8 nm is chosen for the match. The '100 nm' is the corresponding FWHM of a Gaussian function with the standard deviation of 42.8 nm. As is shown in Fig. 5(c), the five layers of the letter-shaped structures are reconstructed at their original depths.

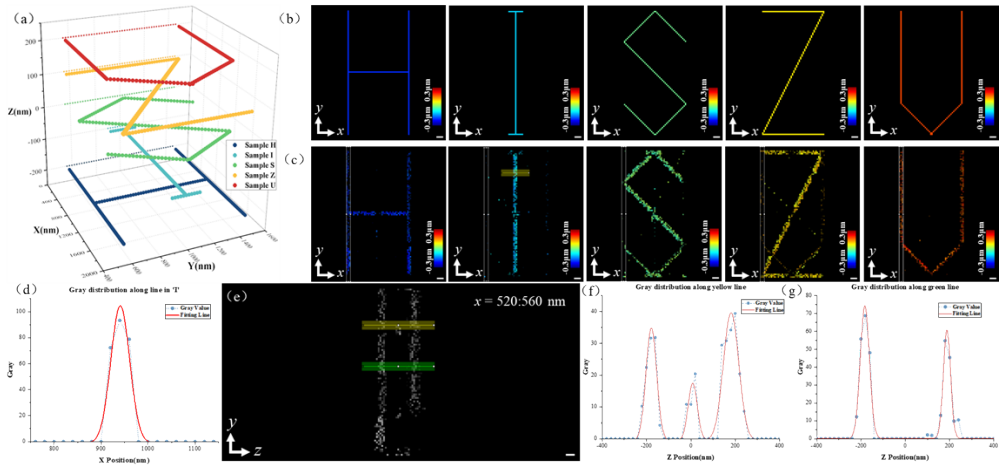

Fig. 5. SH-CS performance with simulated sample. (a) 3D structure of simulated sample; (b) Z-axis projection of the five-letter structure; (c) Reconstructed Z-axis projection of the five-letter structure; (d) Intensity distribution along the yellow line in the second sub-image in Fig. 5 (c) and its Gaussian fit; (e) $yoz$ projection of the white cuboid in Fig. 5 (c); (f) Intensity distribution along the yellow line in Fig. 5 (e) and its Gaussian fit; (g) Intensity distribution along the green line in Fig. 5 (e) and its Gaussian fit. In Fig. 5 (b) and (c), color bar represents the axial depth corresponding to the pseudo-color. Scale bar: 100 nm.

Next, we analyzed the intensity distribution along the yellow line in the second sub-image in Fig. 5(c), and the results are shown in Fig. 5(d). Gaussian fitting of the distribution shows that the FWHM is 54 nm. It is consistent with the expectation that the lateral resolution is which is corresponding to the size of the light needle. In order to analyze the axial resolution, the reconstructed 3D structure within the interval of $x = [520, 560]$ nm was extracted and project it to $yoz$ plane, as shown in Fig. 5(e). The region corresponding to the selected interval is highlighted with white cuboid in Fig. 5(c). The intensity distribution of the yellow line and the green line is shown in Fig. 5(f) ~ (g). The Gaussian function is fitted to their intensity distributions respectively, and the standard deviation of the Gaussian function is about 25 ~ 35 nm.

It should be noted that, in the SH-CS method, off-axis emitter will lead to a lateral displacement of the PSF in the image, and this lateral displacement may be mis-identified as a change in z coordinate. That is why a light needle with confined size is a necessary requirement for the SH-CS. Because only those emitters within the light needle can be excited, the maximum lateral displacement is limited. Furthermore, thanks to Gaussian distribution of the excitation intensity of the light needle, although the emitter with larger extent of off-axis has larger lateral displacement, its fluorescence intensity is lower as well. So, the effect of such lateral displacements is limited. As is shown in Fig. 5 (and Fig.6 in the next section), the effect of the lateral displacement is acceptable for an optical needle of 50 nm (FWHM). If the width of the light needle can be reduced further, the effect of such lateral displacement will be reduced as well.

## 4. Experimental validation

To analyze the performance of the SH-CS in processing of experimental data, we designed a 3D sample (supplementary material for details) composed of fluorescent beads, as is shown in Fig. 6(a). The sample was excited by a light needle (FWHM = 50 nm) which scanned the sample with a step size of 20 nm. A series of raw images were acquired (see supplementary materials for details). Fig. 6(b) shows several raw images obtained under light needle excitation. Only a part of the raw images is shown here, the complete series of the raw images are supplied in supplementary materials (Visualization 1). All raw images are analyzed with the CS algorithm introduced in section 2, and measurement matrix used is based on SH-PSF image of a fluorescent bead at 201 depths in a depth range of 4 μm. The 3D structure reconstructed by SH-CS is shown in Fig. 6(c). As is shown in Fig. 6(c) and Visualization 2 in supplementary materials, the 3D structure of the sample is well recovered. In order to show the axial localizing ability, we analyzed a part of the sample with the densest structure and its reconstructed result, which are highlighted with two rectangles in Fig. 6(a) and Fig. 6(c) respectively. The structures in the rectangle of Fig.6(a) and Fig.6(c) are projected on the *yoz* plane and shown as the upper and the lower sub-figure in Fig.6(d) respectively. The intensity distribution along the green stripe in the lower sub-figure is show as dots in Fig. 6(e), and each peak was fitted with a Gaussian function and shown with green solid line in Fig. 6(e). For comparison, the axial positions of the beads in the corresponding part of the sample are shown as red crosses in the Fig. 6(e) as well. As is shown in Fig.6(e), axial information of some beads is recovered very well, such as the second and the sixth peak, showing high coincidence with positions of their corresponding red crosses. However, other peaks exhibit different degree of offsets or even mis-matching. Since the local density is quite high here, i.e., there are 8 beads within a depth range of ~800 nm, we think the performance is reasonable.

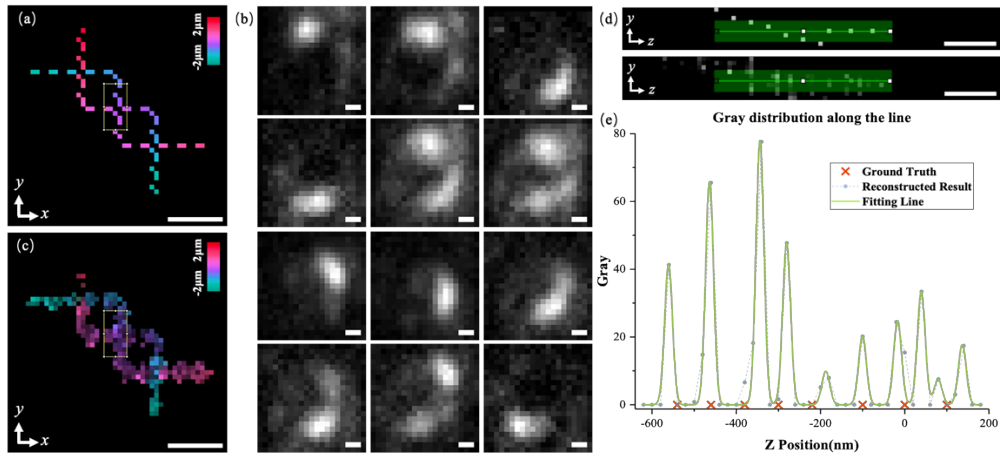

Fig. 6. Using SH-CS to deal with experimental data. (a) Z-axis projection of 3D experimental sample designed by experimental data; (b) Multiple original images obtained by light needle scanning; (c) Z-axis projection of reconstructed 3D sample structure; (d) The structures in the rectangles in (a) and (c) are projected onto the *yoz* plane and shown in the upper and lower sub-figure respectively; (e) Intensity distribution along the green stripe in Fig. 6(d). Color bars in (a) and (c) represent the axial depth corresponding to the pseudo-color. Scale bar: 240 nm.

It should be noted that, the raw images above are not real experimental images acquired directly by a detector in an implemented light needle system, but generated based on experimental images of multiple single beads (see supplementary material for details). Some factors may affect the performance when working with real experiment images. First, the analysis above did not take field-dependent aberrations into account. We believe for real experimental images, the performance of the SH-CS might be improved by building specific measurement matrix for each light needle position, which may be helpful to avoid the field-dependent variation. Second, real images may suffer from imperfect light needle which might weaken the performance of the SH-CS method.

## 5. Conclusion

In order to realize 3D nano-imaging at large depth of field, a new method of recovering axial information of axial dense emitters in a restrained lateral area is put forward here, based on SH-PSF detection and CS algorithm, and it is abbreviated as the SH-CS method. With SH-PSF detection, axial position information is encoded to azimuth angle, so axial dense emitters at different depths can be imaged at different azimuths simultaneously. Then such axial information is decoded using CS algorithm. Simulation data show that SH-CS can distinguish 1 ~ 15 emitters randomly distributed within a depth range of 4 μm with axial localizing accuracy of 12.1 nm ~ 73.5 nm and executing time of 1.1 s. Combined with light needle excitation which can be achieved with BB-STED [10 - 12], the SH-CS was validated with simulated samples or experimental samples. It should be noted that, the lateral resolution is limited by the size of the light needle generated by the BB-STED, which is set to 50 nm (FWHM) here, so if the size is further reduced by using higher depletion intensity, not only the lateral resolution can be improved directly, the axial resolution will also benefit because smaller light needle means lower excited density and smaller possible lateral displacement of off-axis molecules. SH-CS can achieve volumetric imaging method by a 2D scanning. Generally, the larger the imaging area is, the longer the imaging time will be, but this shortcoming will not become a big issue by introducing multi-light-needle parallel excitation, and similar strategy has been used in pSTED [22, 23] and pRESOLFT [24]. For example, pRESOLFT simultaneously uses 100,000 near-ring spots arranged in an array as off-switching light, and achieves ultra-high-resolution imaging in living cells with a temporal resolution of about 1 s [24], but pRESOLFT adopts a confocal detection strategy which has an axial resolution of only 580 nm, and there is only one layer can be recovered at a time, and the volume imaging speed is still limited. So, actually, the SH-CS method can become a complementary tool for the pRESOLFT and pSTED. As to the reconstruction of the SH-CS, although ~1 second is needed for processing one single image so far, however for multiple images acquired, parallel processing will be helpful to improve the total processing time in the future.

**Acknowledgments.** This work was supported by National Key Research and Development Program of China (2022YFF0712500), National Natural Science Foundation of China (Grant Nos. 11774242, 62175166, 61335001), Shenzhen Science and Technology Planning Project (Grant No. JCYJ20210324094200001, JCYJ20200109105411133), Shenzhen Key Laboratory of Photonics and Biophotonics (ZDSYS20210623092006020).

**Data availability.** Data and code underlying the results presented in this paper can be obtained from the authors upon reasonable request.

**Supplemental document.** See Supplement 1 for supporting content.